\documentclass[conference]{IEEEtran}
\IEEEoverridecommandlockouts
\usepackage{cite}
\usepackage{amsmath,amssymb,amsfonts}
\usepackage{algorithmic}
\usepackage{graphicx}
\usepackage{textcomp}
\usepackage{xcolor}
\usepackage{paralist}
\usepackage{enumitem}
\usepackage{amsmath}
\usepackage{subcaption}
\usepackage{booktabs}
\usepackage{wrapfig}
\usepackage[
  draft,
  bookmarks=false
]{hyperref}

\newcommand{\ih}[1]{\textcolor{blue}{\footnotesize [IH: #1]}}%chktex 13
\newcommand{\jm}[1]{\textcolor{orange}{\footnotesize [JM: #1]}}%chktex 13

\def\BibTeX{{\rm B\kern-.05em{\sc i\kern-.025em b}\kern-.08em
    T\kern-.1667em\lower.7ex\hbox{E}\kern-.125emX}}
\begin{document}

\title{Proof-of-Social-Capital: A Consensus Protocol Replacing Stake for Social Capital}

%\author{Regular Paper}
%\iffalse
\author{\IEEEauthorblockN{1\textsuperscript{st} Juraj Mariani}
\IEEEauthorblockA{\textit{Faculty of Information Technology} \\
\textit{Brno University of Technology}\\
Brno, Czechia \\
imariani@fit.vut.cz}
\and
\IEEEauthorblockN{2\textsuperscript{nd} Ivan Homoliak}
\IEEEauthorblockA{\textit{Faculty of Information Technology} \\
\textit{Brno University of Technology}\\
Brno, Czechia}
\IEEEauthorblockA{\textit{Faculty of Informatics and Information Technologies} \\
\textit{Slovak University of Technology in Bratislava}\\
Bratislava, Slovakia \\
homoliak@fit.vut.cz}
}
%\fi

\maketitle

\begin{abstract}
Consensus protocols used today in blockchains mostly rely on scarce resources such as computational power or financial stake, favoring wealthy individuals due to a high entry barrier.
We propose Proof-of-Social-Capital (PoSC), a new consensus protocol fueled by social capital
%\;--\;trust and influence from social interactions\;--\;
as a staking resource to ensure fairness and decentralization.
Consensus nodes in our system do not require financial or computational resources that are expensive to acquire; instead, they utilize social media influence, distributing consensus power not according to wealth but social capital.
Our approach integrates zkSNARK proofs, verifiable credentials with a uniqueness-enforcing mechanism to prevent Sybil attacks, and the incentive scheme that rewards content creators for securing the consensus and optionally followers for engagement with social media content.
%The proposed framework enhances privacy and equity.
This work presents a new concept aligned with the modern social media lifestyle, as applied in finance, and proposes a practical step in the evolution of decentralized consensus protocols.
\end{abstract}

%\begin{IEEEkeywords}
%component, formatting, style, styling, insert
%\end{IEEEkeywords}

\section{Introduction}
% Most of contemporary blockchain consensus protocols, such as Proof-of-Work (PoW)~\cite{nakamoto2008bitcoin,BTCpow} and Proof-of-Stake (PoS)~\cite{casper,ETHwhitepaper,PoSouroboros}, prioritize security but often compromise on privacy or energy efficiency.

%\ih{Ensure that throughout the paper we consider only service-based rewarding of followers. Exception is discussion.}

%The Proof-of-Work (PoW) mechanism utilized by Bitcoin~\cite{nakamoto2008bitcoin,BTCpow} represents the first generation of widely adopted consensus protocols for permissionless blockchains.
%However, PoW raises concerns about energy waste and centralization due to the evolution of specialized (and expensive) mining hardware, which creates a high entry barrier.
The Proof-of-Work (PoW) mechanism used by Bitcoin~\cite{nakamoto2008bitcoin,BTCpow} introduced the first widely adopted permissionless consensus protocol. 
Its reliance on specialized mining hardware raises concerns about energy waste and centralization through high entry costs. 
Proof-of-Stake (PoS)~\cite{casper,ETHwhitepaper,PoSouroboros}, mainly adopted in Ethereum 2.0~\cite{LeastAuthority2018Eth2Audit,casper}, represents the next generation of widely adopted consensus protocols, with the primary goal of eliminating the excessive energy consumption associated with PoW.
However, PoS also favors participants with substantial financial resources, thereby marginalizing those with limited financial capital, leading to the centralization of consensus power and wealth. 

These developments raise a more fundamental question. If we want open participation and resistance to plutocracy in public blockchains, should consensus power be linked to financial wealth? 
PoS removes the energy costs of PoW, but it inherits and worsens the inequality of current wealth distributions. 
Participants who stake little capital remain effectively excluded from securing the protocol, even if they might be key players in an application-level ecosystem. 
We therefore ask whether a different scarce resource, one based on individuals rather than capital, and that is hard to acquire in bulk, can be used to secure consensus.

Social capital captures the value of human capital~\cite{humanCapital}, the attention economy~\cite{attentionEcon}, and social influence metrics~\cite{influenceMetrics}.
Digital platforms monetize attention through follower engagement, illustrating how skills, time, and influence generate measurable economic value.

Social capital has already been applied as a powerful means for value creation in both Web2 and Web3 environments. 
Web2 platforms such as YouTube~\cite{youtube,youtube2}, Instagram~\cite{instagram,instagramTiktok}, and TikTok~\cite{instagramTiktok} operationalize social capital through engagement-driven monetization, but keep substantial revenue shares\footnote{E.g., 55\% on YouTube but varying across platforms~\cite{videoweek}.}~\cite{Hödl2023} and leave creators dependent on supplementary sponsorships~\cite{Hua_2022}.
Web3 platforms like \textit{Farcaster}~\cite{farcaster,farcaster2}, \textit{SteemIt}~\cite{steemwp}, and \textit{Friend.Tech}~\cite{friendtech} integrate social capital at the application layer yet face limited adoption~\cite{frogsanon}, limiting their decentralized and monetization potentials.

Thus, existing designs treat social capital as an aspect of the application-layer while consensus remains tied to stake, separating those who generate engagement from those who have influence over the ledger. 
Our approach instead embeds social capital directly into the consensus layer, so users driving engagement and economic activity also influence block production.
Social capital has not yet been utilized as a means to secure a consensus protocol, which further motivates our work. 

\subsection{Our Approach}
We propose a privacy-preserving consensus protocol that uses social capital as a non-tradable staking resource.
Users allocate social capital ``tokens'' to content creators, which determines the creators' probability of being selected as block proposers.
To ensure privacy and resistance towards Sybil attacks, we combine zkSNARK proofs~\cite{zkp}, verifiable credentials (VCs)~\cite{W3C2019VC}, and a built-in uniqueness mechanism that replaces Ethereum’s global state Merkle Patricia Trie (MPT) keys to embed user identity fingerprints.
By integrating social capital into the consensus layer, our protocol addresses the shortcomings of centralized Web2 revenue models and the limited adoption of Web3 social applications.
It allows creators to earn rewards roughly proportional to their social influence while preserving decentralization and user privacy.
Using real-world YouTube data (see \autoref{sec:implementation:experiments}), we show that our social capital scaling notably reduces inequality in consensus power relative to raw follower counts, mitigating the ``rich get richer'' effect of PoS.
The protocol also lets creators reward their followers with exclusive content as compensation for engagement.

\iffalse
\subsubsection*{\textbf{Organization.}}
Our paper is structured as follows.
\autoref{sec:design} describes the proposed Proof-of-Social-Capital (PoSC) protocol, detailing its entities, mechanisms, processes, and procedures.
\autoref{sec:security-analysis} provides a security analysis, identifying potential attack vectors such as Sybil identities and social capital manipulation, along with proposed mitigation strategies.
\autoref{sec:implementation} outlines the protocol’s implementation details. 
\autoref{sec:discussion} discusses challenges and fields of future work.
\autoref{sec:related-work} reviews related work, comparing centralized and decentralized platforms that leverage social capital.
Finally, \autoref{sec:conclusion} concludes the paper with a summary of the proposed protocol.
\fi
\section{Problem Definition}
\label{sec:problem}

% \subsubsection{\textbf{Problem:}} 
We intend to create an alternative type of consensus protocol based on social capital that secures the blockchain and replaces traditional resources, such as stake, storage, or work.
We thus consider social capital as a non-trivial resource to acquire (similar to other mentioned resources) and should be explicitly manifested on the blockchain platform through its users.
Nevertheless, designing such a system has some challenges, such as mitigating overly large individuals, ensuring the uniqueness of all users within the system, and, at the same time, preserving their privacy, which is the focus of our work.

\subsection{Attacker Model}
%\ih{popisat problem, motivacia, assumptions - trusted issuers, secure cryptography primitives, max 0.5page, entity, podla vyoru etherea, byzantine odolnost, nody si neveria, ... , podsekcia attacker model, pozri clanok smart otps}
We assume the goals of the attacker are as follows:
\begin{compactenum}
    \item Create Sybil entities to artificially increase the social capital of the attacker. % and profits from the consensus protocol. %bias the leader election results in their favor.
    
    \item Misuse of dominance of the consensus power due to immense social capital. % by: monopolizing social capital and coercion, collusion, or other off-chain practice.

    \item Reveal the identities of other users (from data stored on blockchain). % via:
    % \begin{inparaenum}    
    %     \item identity verification processes,
    %     \item identities stored on-chain, and
    %     \item capturing IP addresses.
    % \end{inparaenum}
    
    %\item Prevent other validators from producing blocks (tangential goal).
\end{compactenum} 

\smallskip \noindent
We assume that standard cryptographic primitives cannot be broken by polyno\-mial-time adversaries. 
In particular, asymmetric encryption and signature sche\-mes are assumed to provide semantic security and existential unforgeability, respectively. 
The employed cryptographically secure hash function $H(.)$ is modeled as a random oracle, ensuring collision resistance, preimage resistance, and second-preimage resistance.
Furthermore, the utilized zkSNARKs are assumed to be sound and zero-knowledge, i.e., adversaries cannot generate valid proofs for false statements nor extract hidden witness information from valid proofs. 
We also assume that all third-party identity providers (e.g., issuers of verifiable credentials) are trusted, which is a common assumption for systems building on external identity~\cite{GrunerEtAl2018_RelevanceBlockchainIdentity,Hyperledger_Weaver_DezIdentity_2024,HardjonoPentland2016_PrivacyPreservingIdentity}. % \ih{JM: citacie}. % to change their operation by an attacker.
Finally, we assume aspects of the existing Ethereum PoS consensus protocol, which we modify in terms of replacing financial stake with social capital. %\ih{Add assumption about only a single PoSC platform. - Maybe put into the convenient place in the discussion.}

\section{Proposed Approach}
\label{sec:design}

\begin{figure}[t]
%    \vspace{-0.4cm}
    \centering
    \includegraphics[width=0.5\textwidth]{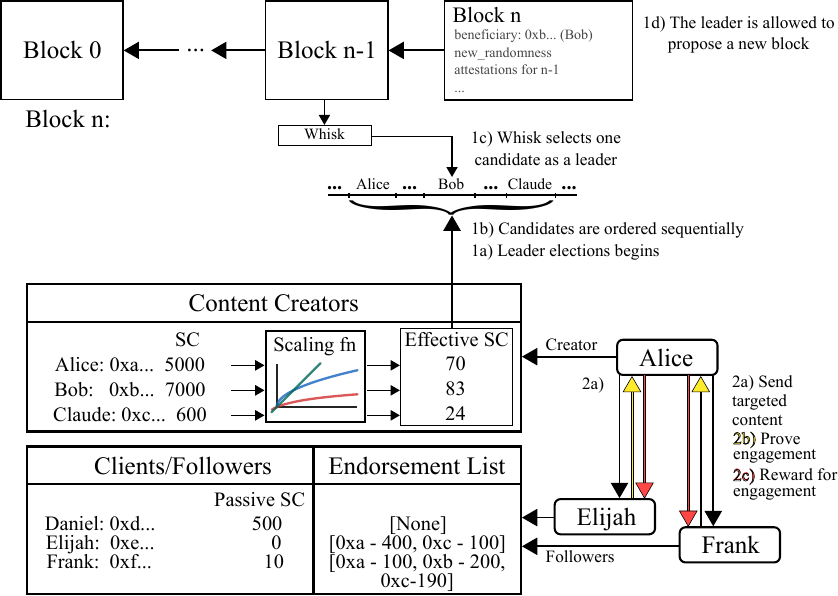}
    %\vspace{-0.2cm}
    \caption{Overview of the proposed protocol. Rewarding scheme for followers is optional.}%\jm{dat reward ako optional}} 
    \label{fig:design:overview}
    %\vspace{-1em}
\end{figure}

In this work, we propose a Proof-of-Social-Capital (PoSC) consensus protocol that works in a similar fashion as Ethereum's Proof-of-Stake protocol~\cite{ethereum-pos-mechanism}, but replaces the stake with social capital (see overview in \autoref{fig:design:overview}).
Content creators are endorsed by their fans, receiving their social capital. Based on received social capital, they could be elected as block proposers, with a probability almost\footnote{To avoid large monopolies, social capital is scaled (see \autoref{sec:design:overview:scap}).} proportional to their social capital.
Therefore, content creators can produce blocks and earn block rewards (steps 1x in \autoref{fig:design:bootstrapping:idVerification}), while followers can be, optionally,  rewarded for receiving targeted content from content creators (steps 2x in \autoref{fig:design:bootstrapping:idVerification}). 
%\jm{better argument for why PoSC would be better than PoS.}

%Content creators accumulate social capital through endorsements from followers.
% Nevertheless, if the system is designed naively, Sybil entities might emerge and artificially award the social capital of Sybil accounts to themselves. 
% Therefore, we have to avoid Sybil entities and ensure the uniqueness of users within the system.
%We detail our system in the following.
\iffalse
    \footnote{Note that for simplicity, in the context of this work, we will talk about `system' instead of `platform' since we propose a proof-of-concept design. However, in the future, our consensus protocol might provide smart contract capability and thus act as a platform for other smart contracts.} 
\fi

\subsection{System Entities}

The system comprises two distinct types of entities:

\begin{compactitem}
	\item \textbf{Consensus nodes (i.e., content creators)}: users with high enough social influence.  
	They accumulate social capital through endorsements from regular users (who act as followers) and stake it to join the consensus.  
	Their consensus power is proportional to the support received from followers.
	
	\item \textbf{Lightweight clients (i.e., followers)}: users who receive and interact with the content. 
    They have a limited (passive) social capital budget assigned, which they allocate to consensus nodes through endorsements (making the social capital active).  
	Thus, they influence the distribution of consensus power.  
	% They may also operate as full nodes, maintaining the ledger and verifying transactions, but they do not take part in leader election or block production.
\end{compactitem}

\subsection{User Uniqueness and Privacy}
\label{sec:design:overview:id&priv}

Since followers manually endorse content creators, a naive design would enable attackers to create a large number of Sybil followers and inflate their social capital.
Therefore, partial identity verification is required to enforce a one-account-per-person policy and ensure that the system contains only legitimate social capital.
We consider verifiable credentials (VC)~\cite{W3C2019VC} signed by any of the trusted authorities, which should contain (mostly private) information about the user's identity, such as name, surname, permanent residence address, birth number, place of birth, nationality, ID of VC, issuance and expiration dates, etc.,
as well as and the user's public key to our PoSC blockchain platform ($PK_B^u$).
All of the mentioned fields 
but $PK_B^u$ 
are concatenated and hashed as \textit{ID-hash}, %\ih{check that it is mentioned first time here}
which serves as a public identifier of user identity on the blockchain as well as the key to the global state of the blockchain -- i.e., sparse Merkle Patricia Trie (MPT).\footnote{Note that all input fields must be extracted from a VC and it should have enough unpredictable entropy to avoid bruteforcing attacks targeting privacy of user data.
We estimated the entropy of these fields to be within $256$ and $530$ bits (assuming 32B for ID of VC), and realistically around $314$ bits.
Even if we assume the worst-case scenario (only the ID of VC is unknown), it still provides a minimum of $256$ bits of entropy.} 
Therefore, the uniqueness of identity is ensured by utilizing MPT with radix 16, as in Ethereum.  
If we were to assume that our system would support 10 billion users, then a full MPT would fit within $log_{16}(10^{10}) \approx 8.3$ levels, which is comparable to Ethereum's circa 325 million accounts~\cite{ethNoAddr}: $log_{16}(325M) \approx 7$.
%\ih{Expand current impact on the size of Ethereum (325M accounts with added 32B od Id-hash and 288B of zkSNARK proof + $\approx$1.75x for overhead.)}

\subsection{Updating Global State}
% ID-hash entries are inserted into the global state upon successful identity verification.
%
The identity verification (see also \autoref{sec:design:enrollment}) is made on-chain in the precompiled \textit{identity-verifying smart contract} ($SC_{ID}$) that inserts ID-hash into the global state, and it encompasses the following checks:
\begin{compactenum}
    \item A VC must be issued by one of the trusted issuers listed in $SC_{ID}$ (that is managed by a majority consensus of nodes).

    \item ID-hash does not exist in the global state yet (i.e., identity uniqueness).

    \item A registering user provided a valid zkSNARK proof~\cite{groth2016} with private parameters corresponding to all VC fields but $PK_B^u$, where the following holds:
    \begin{compactitem}
        \item Ownership of VC: registering tx was signed by the private key corresponding to $PK_B^u$ in VC.

        \item ID-hash = $H(VC.\{private\ identity\ fields\})$.
    \end{compactitem}
\end{compactenum}

% To preserve privacy (of personal data in a VC), our system uses zkSNARKs~\cite{groth2016} that allow users to prove ownership of their verifiable credentials (VC)~\cite{W3C2019VC} 

% and their uniqueness w.r.t. existing users in the system.  
%
% We utilize Groth16~\cite{groth2016}, widely adopted in Ethereum due to its small proof size and efficient verification, making integration practical.
% Alternatives and impacts are discussed in \autoref{sec:security-analysis:zkps} and \autoref{sec:discussion:pqc}.
%

The zkSNARK proofs are not stored within the $SC_{ID}$ smart contract and are only passed as calldata within the registering transaction, meaning that additional storage requirements of $SC_{ID}$ are only the data overhead of the MPT, along with the metadata stored in the leaf nodes (i.e., VC expiration date, etc.).
As a result, all identities remain private by default, although some users may opt for publicity (i.e., content creators).  
In particular, to obtain sufficient social capital, content creators publish their PoSC blockchain address in their Web2 social network profiles, which are visible to followers and the public.

\subsection{Social Capital: Types and Scaling}
\label{sec:design:overview:scap}
%\ih{minimalny socialny kapital pre consenus nody 10000 nasobok maxima pasivneho kapitalu pod+la iteracii pri elections}
Social capital is divided into passive and active.
The \textbf{\textit{passive social capital}} budget assigned to every unique user of the system can be awarded to content creators by endorsements.
After awarding, it becomes  \textbf{\textit{active social capital}}, reflecting the consensus power of the content creator. 
%whereas regular users (i.e., followers) possess only passive social capital.
%
% In the context of our work, we define \textbf{\textit{social capital}} as the amount of \textbf{passive social capital} a creator has accumulated.
Since every user with a unique identity is assigned a fixed amount of passive social capital budget, it effectively becomes a scarce resource.
% Active social capital reflects the consensus power of the content creator.

To prevent the accumulation of excessive consensus power by certain highly influential content creators, we propose %\textbf{\textit{dynamic}}
scaling of social capital that mitigates this problem, and thus improves decentralization (see below).  
% This limits the ability of one or more highly endorsed participants to dominate the network and strengthens the system’s resistance to centralization.  
We refer to the output of the scaling as \textbf{\textit{effective social capital}}, which directly determines the consensus power of a node (see also \autoref{sec:security-analysis:social-capital:monopoly}).
%\ih{specify that opportunistic assignment of social capital due to lucrative creators is partially combatted by the diminishing returns of the scaling function. As assigning more social capital does not linearly increase the chance of becomming a leader, such assignments are not as valuable as would one think.}
% \jm{We could also implement a vote obfuscation technique?}

%
% On the other hand, this approach ensures that even nodes with lower social capital have a reasonable chance of producing blocks. 
% ...while highly endorsed nodes have a reduced probability\footnote{Amassing large amounts of active social capital does not translate linearly to consensus power.} 
% of being selected as leaders—thus encouraging a more decentralized consensus process and countering monopolization attempts. 
% The impact of the scaling function and the use of effective social capital can be seen demonstrated in \autoref{sec:security-analysis:social-capital:monopoly}.

% An argument to a question regarding protocol scalability related to the number of possible content creator nodes, we propose a limit to the electable nodes. This limit is explained later, in \autoref{sec:design:bstrap:initialization}. \ih{nemusime o tom hovorit uz teraz}

\subsubsection{\textbf{Scaling Social Capital}}
The concept of effective social capital should help to balance the effective distribution of social capital through scaling.
We consider two possible options for the scaling effect: (1) \textit{quadratic voting} and (2) a \textit{non-linearly decreasing scaling function} (such as $sqrt()$ and $log_2()$).
Quadratic voting~\cite{posner2015voting,qvote} scales down the input voting power before assignment to the content creator, which preserves the relative ratios of obtained votes (i.e., active social capital) for all content creators, and thus does not help to resolve the problem (see also \autoref{tab:security-analysis:social-capital:monopoly}).\footnote{Moreover, quadratic voting decays the weight of the multiple divergent votes made by a single follower, while a vote for a single choice has the highest weight.}
On the other hand, the considered scaling functions scale down the obtained number of votes post-assignment, which causes a desired change in the relative ratios of such scaled-down votes among content creators.

%
\iffalse
For example, if two voters with $100$ voting credits were to vote for the same option with their maximum voting power, the voting power of the selected choice would be $\sqrt{100}+\sqrt{100}=20$.
In quadratic voting, the voting power of the selected option would be $\sqrt{100}+\sqrt{100}=20$ in contrast to $\sqrt{100+100}\approx14$ of the square-root scaling function.
We chose the non-linear scaling function due to its simplicity and adjustability. \ih{improve reasons.}
\fi

% Thus, we propose the following mechanisms.

% % function (e.g., $log(.)$ or $sqrt(.)$) 

% \begin{compactitem}
%     \item Quadratic voting
%     \item (Square root) Scaling function
% \end{compactitem}
%

\subsection{Leader Election and Consensus Power}
\label{sec:design:overview:leadere-cpower}

Validators "stake" their active social capital (becoming effective social capital) to participate in the leader election process. % and, consequently, in block creation.  
Our protocol takes inspiration from Ethereum’s Randao~\cite{LeastAuthority2018Eth2Audit} and WHISK-SSLE~\cite{whisk} family of protocols%(see also \autoref{sec:security-analysis:consensus:election})
, adapting their randomness-based leader election mechanisms to generate on-chain randomness with Randao, and elect block producers in an unpredictable way using Whisk to hinder leader prediction (and subsequent DoS) attacks possible with Randao (nonetheless, it remains an ancillary aim).
To reduce the risk of inter-round stake lobbying, we incorporate Ethereum’s delayed staking mechanism~\cite{LeastAuthority2018Eth2Audit}. % \ih{ref}.  
With this approach, when a node requests to join the validator pool, it must wait a certain number of blocks before joining. %(see also \autoref{sec:security-analysis:consensus:election}). % can begin participating in the protocol.  
%This delay increases the difficulty of exploiting the randomness source and lowers the chances of successful attacks against the Randao mechanism (further details can also be found in \autoref{sec:security-analysis:consensus:election}). \ih{nie je potrebne}

\subsection{Incentive Scheme}
\label{sec:design:overview:incentives}

Validators (i.e., content creators) are incentivized with native token rewards for valid block production and penalized through slashing for protocol violations (e.g., inactivity, equivocation, invalid blocks, etc.). 
Followers might benefit economically from endorsing creators, who may unlock exclusive content or reward followers for engagement with the common content (e.g., ads or campaigns).

Since social capital serves as an identity-related consensus resource, followers should assign it to content creators without paying transaction fees.
Therefore, creators cover fees via delegated meta-transactions (i.e., ERC-4337~\cite{ERC4337}). % preventing spam while giving creators discretion to process endorsements.
\subsubsection{\textbf{Penalization}}\label{sec:design:overview:incentives:penality} To enhance the protocol's safety (and finality), validators attest every block. 
Failures or malicious actions should be penalized.  % slashing of both tokens and social capital.
We propose several options for penalization:
\begin{compactenum}
    \item Return of social capital to original endorsers and temporarily banning offenders from consensus and receiving endorsements, with severity-based scaling.
    \item Adjustment (temporary) of the creator’s scaling function by a multiplicative constant or to a steeper root (e.g., $\sqrt{\text{active SC}}\rightarrow \frac{1}{c}\sqrt{\text{active SC}}$ or $\sqrt[\leftroot{-1}\uproot{2}\scriptstyle 3]{\text{active SC}}$).
    \item Increasing endorsement fees, i.e., fees for accepting social capital.
\end{compactenum}
\par\noindent
We adopt the second penalization strategy: \textit{modifying the creator's scaling function} (either via a multiplicative constant or by switching to a steeper root) because it imposes a non-negligible economic penalty that scales with the creator’s current stake and the severity of the violation while still preserving the total amount of delegated social capital in the system.%\ih{better reason.}

An argument might arise about the effectiveness of such penalizations, predominantly in an environment where a malicious creator could simply transition onto another PoSC platform.
This situation is later discussed in \autoref{sec:discussion:slashing}.
%\ih{also acknowledge or otherwise `solve' the slashing problems (see review suggestions from FC26).}
%\ih{See Section X for discussion about the effect of penalization.}
% it limits collateral damage to the attacker's followers (e.g., endorsement revocation or fee increase).

\iffalse
Extensions may increase penalties by or for offenders \ih{Toto je velmi konceptualne, chcelo by to byt presnejsi kedze popisujeme pristup.}.

\ih{Ten priklad by som vobec neuvadzal. Skus to skor zase vylistovat ako moznosti a vyberieme len jednu -- tu kratenie socialneho kapitalu na urcity pocet kol.}

\subsubsection{\textbf{Example of Post-Slashing Parameter Adjustment}} In a system where Alice is a content creator with an active social capital equal to 1000 units, scaled by a square root function to 31 units of effective social capital.
If her activity constitutes a slashable offence, a static percentual portion of her active social capital (i.e., 10\% for the first offence) will be deducted from her active social capital reserve and returned to the respective followers.\ih{Ktorym followerom?}
Additionally, her scaling function could be dynamically adjusted by a multiplicative constant ($\sqrt{\text{aScap}} \rightarrow \frac{1}{c}\sqrt{\text{aScap}}$) for a predetermined time period (in epochs).
Lastly, the associated fee for accepting endorsements could be (temporarily) increased to discourage reoffending.
\fi

\subsection{User Enrolment}
\label{sec:design:enrollment}

A robust user enrolment ensures that users possess legitimate VCs and are unique within the system.
It follows these steps (see also \autoref{fig:design:bootstrapping:idVerification}):
\begin{figure}[t]
	\centering
	\includegraphics[width=0.55\linewidth]{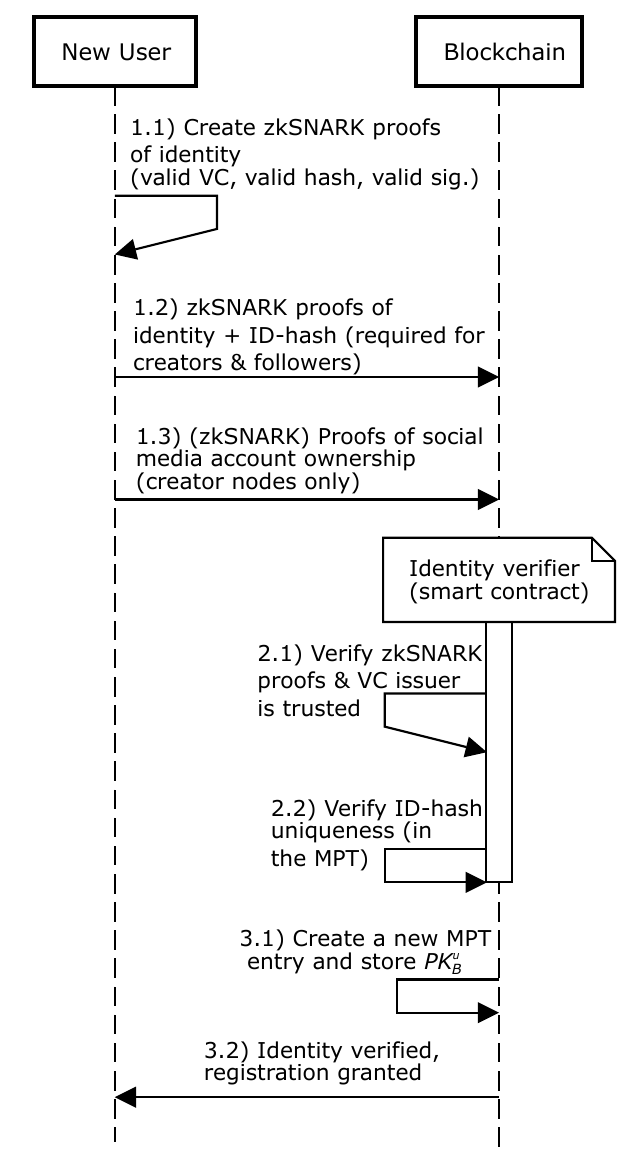}
    %\vspace{-1cm}
	\caption{Identity verification process.%\jm{mensi font} 
    % The verifier could be a smart contract or a part of the consensus logic. A new node can be either a consensus node (content creator) or a client node (follower).
    }
	\label{fig:design:bootstrapping:idVerification}
	%\vspace{-2em}
\end{figure}
\begin{compactenum}
    \setcounter{enumi}{-1}
    \item  \textbf{VC Acquisition}: A user needs to contact a trusted issuer and request a VC.
This action happens off-chain and outside of this system.
\item \textbf{Proving Identity \& Web2-Based Social Capital}: All users create zkSNARK proofs of their identity. % (i.e., followers and content creators) and of social media account ownership (i.e., content creators).
This proof proves the VC the user possesses is genuine (i.e., has a valid issuer).
The proof also testifies to the validity of private personal information in VC data through ID-hash (see \autoref{sec:design:overview:id&priv}).
Additionally, to avoid stealing someone else's social media reputation, a content creator proves the ownership of a social media account.
In detail, a content creator includes their blockchain address on their social media account and constructs a SNARK proof testifying that the account's Web2 page has been signed by the key from the platform's HTTPS certificate.
 \item \textbf{Uniqueness Verification}: 
ID-hash serves to prove the uniqueness of the user's personal information from VC by checking its existence in the global state, which is triggered from $SC_{ID}$.

\end{compactenum}

\subsection{Bootstrapping}

Bootstrapping the consensus protocol involves initializing the network, verifying participant identities, and ensuring the fair distribution of social capital tokens to establish a functional decentralized system.
%
% \subsubsection{\textbf{Staking Social Capital.}}
After verifying their identity, consensus nodes stake their social capital to take part in transaction validation and block proposal. 
% This process involves accessing their default \textit{active social capital} balance.

\subsubsection{\textbf{Initialization}}
\label{sec:design:bstrap:initialization}

We identified two potential methods for determining the initial amount of active\footnote{See \autoref{sec:design:overview:scap}.} social capital:
\begin{compactitem}
	\item Each consensus node starts with the zero value of active social capital and earns it by endorsements from followers.
	
    \item We import social capital from external sources via oracles (either centralized~\cite{zhang2016towncrier} or decentralized~\cite {ellis2017chainlink}). 
    Moreover, if multiple such Web2 social network sources were to be utilized, a mean value would be calculated.
\end{compactitem}
\begin{figure}[t]
	\centering
	\includegraphics[width=0.65\linewidth]{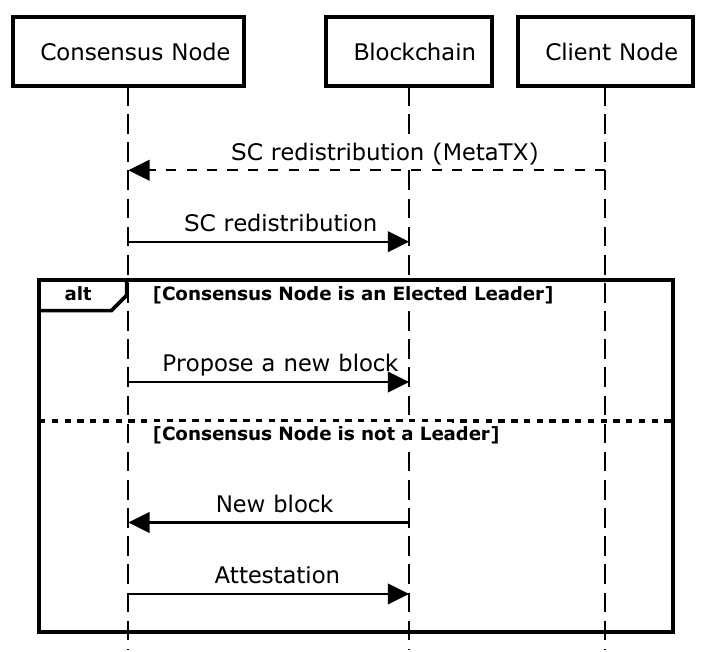}
	\caption{Block proposal and (re)assignment of social capital. The dashed line is used for off-chain messages.}
	% pridat to dvojite podpisovanie distribucie SC
	\label{fig:design:bootstrapping:netFunction}
	%\vspace{-2em}
\end{figure}
%\par\noindent\parshape=0
We opted for the first option since it enables us to maintain social capital distribution on our platform without any external dependencies.
\subsubsection{\textbf{Consensus Threshold}}
%\par\noindent\parshape=0
Currently, active social capital only serves the purpose of staking, and we treat it as automatically staked for all content creators.  
However, an excessively large number of content creators would produce a potentially significant computation overhead in effective social capital calculation (see \autoref{sec:implementation:experiments}).

%
%
%\par\noindent\parshape=0
Thus, a threshold shall be added to limit participation of creator nodes to only those whose number of endorsements received %(i.e., their active social capital) 
exceeds 10,000 times the value of a full endorsement.\footnote{A full endorsement denotes an endorsement awarding all passive social capital to one content creator, instead of distributing it to multiple creators.}
Based on data estimates on the platform YouTube, there are circa 3 million~\cite{Broz2025YouTubeChannelsStatistics,Thumblytics2024YoutubersOver1M} accounts with more than 10,000 subscribers. 
With this \texttt{consensus participation threshold}, we achieve a limit close to the current number of validators in Ethereum of around 1 million~\cite{BeaconchainValidators2025}. %(see also \autoref{sec:implementation:experiments}). 

\subsubsection{\textbf{Block Proposal Mechanism and Social Capital Redistribution}}
\label{sec:design:sc-redistribution}

The protocol employs a leader-based block proposal model akin to those found in Proof-of-Stake systems.
% We modify Ehteruem's v2.0 protocol~\cite{LeastAuthority2018Eth2Audit,casper} 
% however, even other PoS consensus protocols can be utilized.
The leader election is determined by a node’s effective social capital, proportionally to the total effective social capital in the network.
Elected leaders propose new blocks, while non-elected nodes issue attestations (both are rewarded).
We follow the inflationary model of Ethereum due to inheriting its PoS-based attestations.
Upon reaching consensus, blocks are finalized and appended to the ledger.

Social capital balances are updated every round. %based on user activity.
To mitigate strategic manipulation, redistributions of social capital are delayed by a fixed number of epochs. %allowing the system to stabili stabilization.
Since a content creator is endorsed, the follower's social capital remains in the creator's possession unless explicitly reassigned.
We propose a mechanism where a newly assigned creator of social capital will fund the transfer metatransaction of social capital to their address from the original creator. 
% to themselves, if there exists a transfer metatransaction initiated by that follower.

\iffalse
The mechanism aligns with Ethereum’s availability-based consensus model, which prioritizes network availability over immediate consistency.
This design enhances scalability by tolerating partial node inactivity, though at the cost of short-term consistency.
As in Ethereum, blocks remain non-final during an epoch and may be reverted if later deemed invalid, with penalties applied to misbehaving proposers and attestors.
Finality is achieved at the epoch’s conclusion through checkpointing.
\ih{nie je to pre nas podstatne}
\fi

% \subsubsection{\textbf{Social Capital Redistribution.}}
\section{Evaluation}
\label{sec:implementation}
%\ih{podla vtubers clanku skusit aplikovat effective social capital a ukazat zmenu gini indexu, whisk overhead, effective social capital calculation overhead experiment (1000nasobok, 10000x), bud nase experimenty s BLS alebo citovat existujuce experimenty agregacie, specificke experimenty pre PoSC, rozne pocty validatorov, tabulka vysledkov, evalujeme len specificke veci PoSC}\ih{GINI: 0.2886}

\subsection{Implementation}
To evaluate the feasibility of our proposed PoSC consensus mechanism, we developed a proof-of-concept implementation~\cite{scap} in \texttt{Python 3.10} (consisting of around 3,500 LOC), with ZoKrates~\cite{zokrates} used for compiling and verifying PoU zkSNARKs. %\ih{dole nejake merania casu a pamate; verifikacia aj generovanie dokazu}
% Python was selected for its rapid prototyping capabilities and simplicity, while ZoKrates offers a developer-friendly interface to R1CS-based proof systems compatible with Ethereum-style environments.
The system supports core protocol features, including:

\begin{compactitem}
  \item Transactions re\-pre\-sen\-ted as \emph{rlp.Serializable}~\cite{pyrlp} objects,
  \item On-chain identity (ID-hash w. zkSNARKs as described in \autoref{sec:design:enrollment}.) storage using Ethereum Foundation's \verb+HexaryTrie+ structure~\cite{pytrie} for identity-hash uniqueness enforcement, %\ih{bud viac specificky v zmysle pouzitych datovych struktur a algoritmov a kniznic.}
  \item Verification of zkSNARK proofs using ZoKrates CLI environment~\cite{zokrates}, %\ih{kniznice a ref}
  \item Social capital endorsements with (off-chain) delegated meta-transactions for creator-paid social capital assignments. 
  Endorsements are special transactions (with transaction \verb+type+  field equal to \texttt{MetaTx}),
  \item A static square root scaling function of active social capital,%\footnote{\ih{spomenut, ze by as to mohlo dat spravit ako sietovy parameter dynamicky.}}
  \item A Randao-based leader election\footnote{Note that the implementation of Whisk is orthogonal to our proposed protocol and we omitted it in PoC implementation.} and on-chain randomness. 
  Instead of aggregated BLS signatures, \verb+secp256k1 ECDSA+ signatures are collected.
\end{compactitem}
%\ih{The source code of our implementation is available at <anonymous link - dropbox alebo github>}

\medskip \noindent
%\ih{Prepisat aby sme to tak neodstrelovali ale radsej odovodnili preco sa to tak spravilo. Ospravedlnenie, ze sa nam jekdnalo o impl. verifikacie a generovnania zk dokov len ako podcast nasho celkoveho systemu.}
Our implementation enables partial identity verification.
In particular, we implemented a zkSNARK circuit for credential proof generation.
However, given that certain primitives (e.g., BLS signatures) are implemented by 3rd parties, for simplicity and time-saving reasons, we decided to omit them from the proofs.
In their place, traditional \texttt{secp256k1 ECDSA} signatures are used.

The implementation is not focused on full identity verification. % but on as an additional complexity within the system.
%, though the circuit is not currently integrated into the protocol’s execution path; instead, mocked verification values are used for testing purposes.
%Likewise, follower incentive mechanisms\;--\;such as protocol-level tools for rewarding interaction with the targeted (e.g., marketing/influential) content\;--\;have not yet been implemented.
It serves primarily as a validation tool for the protocol’s architectural design and core logic, rather than as a performance-optimized or production-ready deployment.
Moreover, government-issued certificates in eIDs (of EU countries) contain limited fields out of the ones we mention in \autoref{sec:design:overview:id&priv}, which is the subject for their improvement as well as using zkSNARK-friendly cryptographic constructs such as Posseidon~\cite{Grassi2019Poseidon} hashes and BLS~\cite{Boneh2001BLS} or Schnorr~\cite{Schnorr1991Schnorr} signatures.

\subsection{Evaluated Experiments}
\label{sec:implementation:experiments}
%\ih{dolezite}
We conducted experiments mainly to understand and quantify the impacts of the shift from PoS to PoSC. %\ih{zvyraznit PoS problematikou sa nezaoberame, len specificke vlastnosti PoSC}
As we build upon the principles of PoS, some aspects of performance remain mostly the same (e.g., transaction throughput, scalability, etc.).
Therefore, we focus on the added complexity of our PoSC protocol and its main sub-components: overhead of generating and verifying ZK proofs, scaling function application, and the number of validators during leader election.
Moreover, we give an example of how using a PoSC platform could affect income inequalities currently present on social media platforms.
%\ih{spomenut aj gini}

\subsubsection{\textbf{Gini Coefficient and Social Capital}}
\begin{figure*}[t]
    \centering
    % Row 1
    \begin{subfigure}{0.37\textwidth}
        \centering
        \includegraphics[width=1\linewidth]{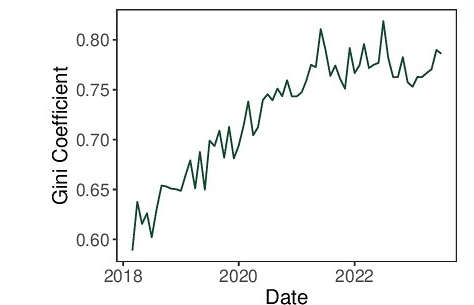}
        \caption{Gini coefficient of current monetization on YouTube of content creators of the VTuber genre~\cite{incomeinequality}.}
        \label{fig:eval:Vtubersgini}
    \end{subfigure}
    \begin{subfigure}{0.37\textwidth}
        \centering
        \includegraphics[width=0.9\linewidth]{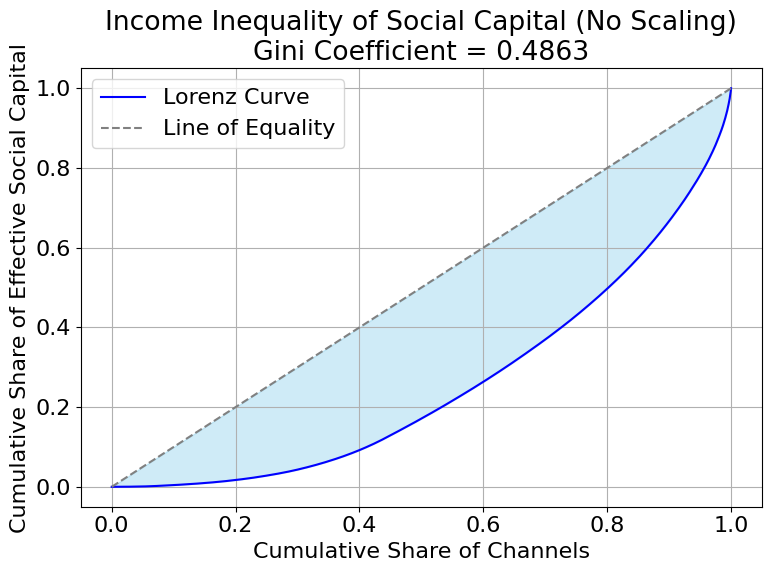}
        \caption{Gini coefficient of raw follower numbers.}
        \label{fig:eval:noscaling}
    \end{subfigure}
    
    % Row 2
    \vspace{1em} % vertical spacing between rows
    \begin{subfigure}{0.37\textwidth}
        \centering
        \includegraphics[width=0.9\linewidth]{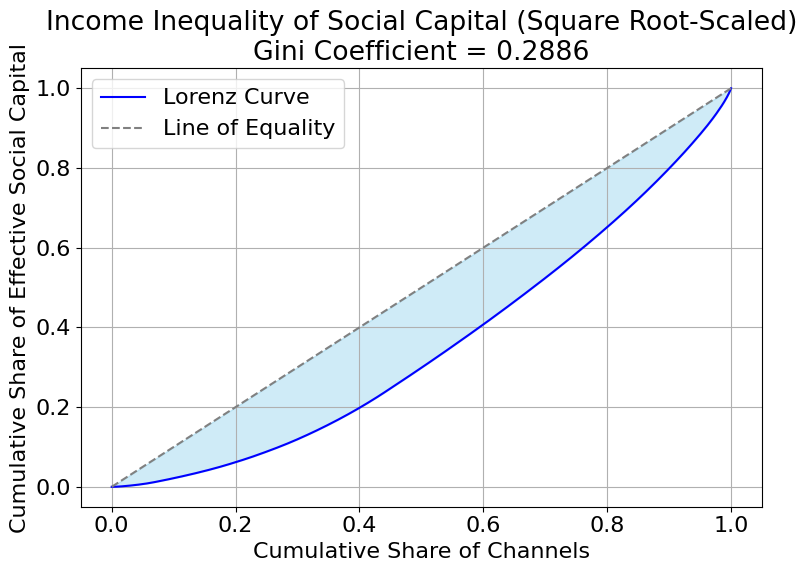}
        \caption{Gini coefficient of social capital with a square root scaling function.}
        \label{fig:eval:scapgini}
    \end{subfigure}
    \begin{subfigure}{0.37\textwidth}
        \centering
        \includegraphics[width=0.9\linewidth]{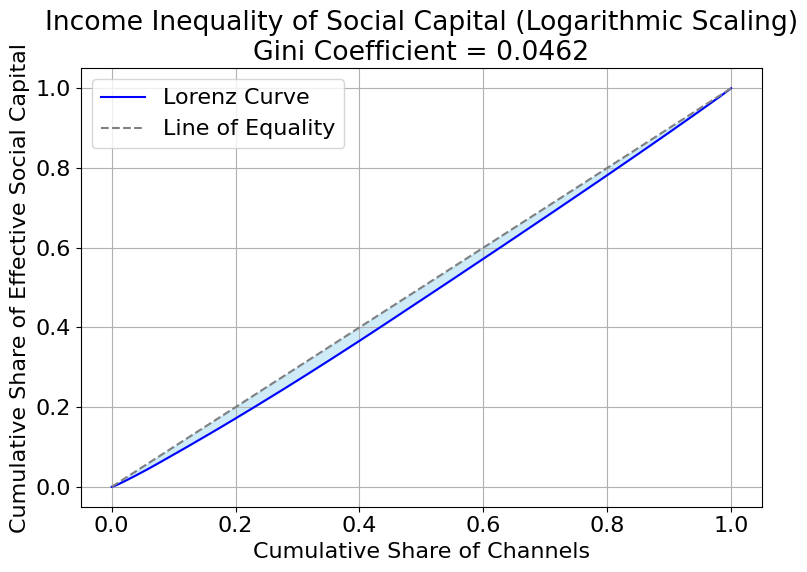}
        \caption{Gini coefficient of logarithmically scaled social capital.}
        \label{fig:eval:logscaling}
    \end{subfigure}
    
    \caption{Gini coefficients in different scenarios of platform monetization.}
    \label{fig:grid}
    \vspace{1em}
\end{figure*}

%\jm{A better dataset of 1M YouTube channels shows a Gini coefficient of an incredible 0.9688. Reflect this change in the article.}
We study how applying our scaling function to social capital impacts Gini inequality within the system.
Our dataset comprises 2,250 YouTube channels and their subscriber counts, which we treat as a proxy for active social capital, without further granularizing endorsements (unlike our main approach or quadratic voting).
To prevent monopolization and give smaller creators a chance, we scale raw follower counts to obtain effective social capital; the scaling functions and their impact on the Gini coefficient are shown in \autoref{fig:grid} (b, c, d).
Without scaling, the Gini coefficient of social capital is $\approx0.49$, indicating substantial inequality (i.e., a small number of highly successful individuals).
Applying a logarithmic scaling function mitigates inequality most strongly (by about $90\%$), although we prefer some inequality to remain.

An even higher income disparity ($\approx0.78$) was reported by \textit{Zhao et al.}~\cite{incomeinequality}, who computed the coefficient for VTuber creators assuming only direct follower support (see \autoref{fig:eval:Vtubersgini}).
Under comparable assumptions, square-root scaling reduces the Gini coefficient from $\approx0.78$ to $\approx0.29$, substantially lessening inequalities in content monetization.

\subsubsection{\textbf{ZKP Overhead}}

\begin{figure*}[t]
    \centering
    \begin{subfigure}{0.49\textwidth}
        \includegraphics[width=\textwidth]{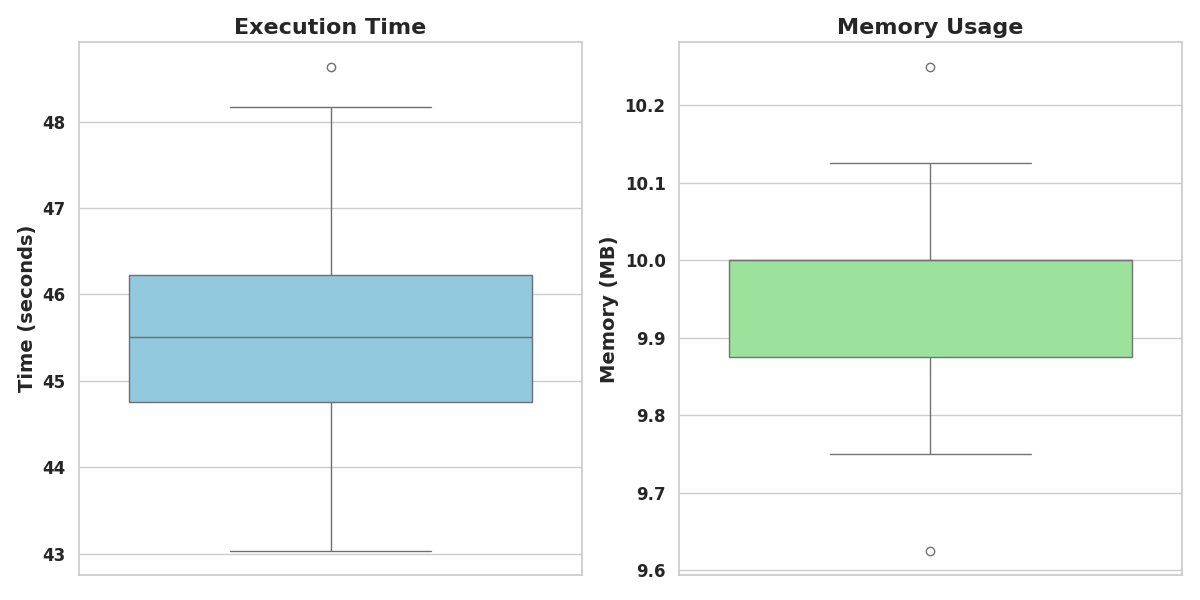}
        \caption{Zk proof generation requirements.}
    \end{subfigure}
    \hfill
    \begin{subfigure}{0.49\textwidth}
        \includegraphics[width=\textwidth]{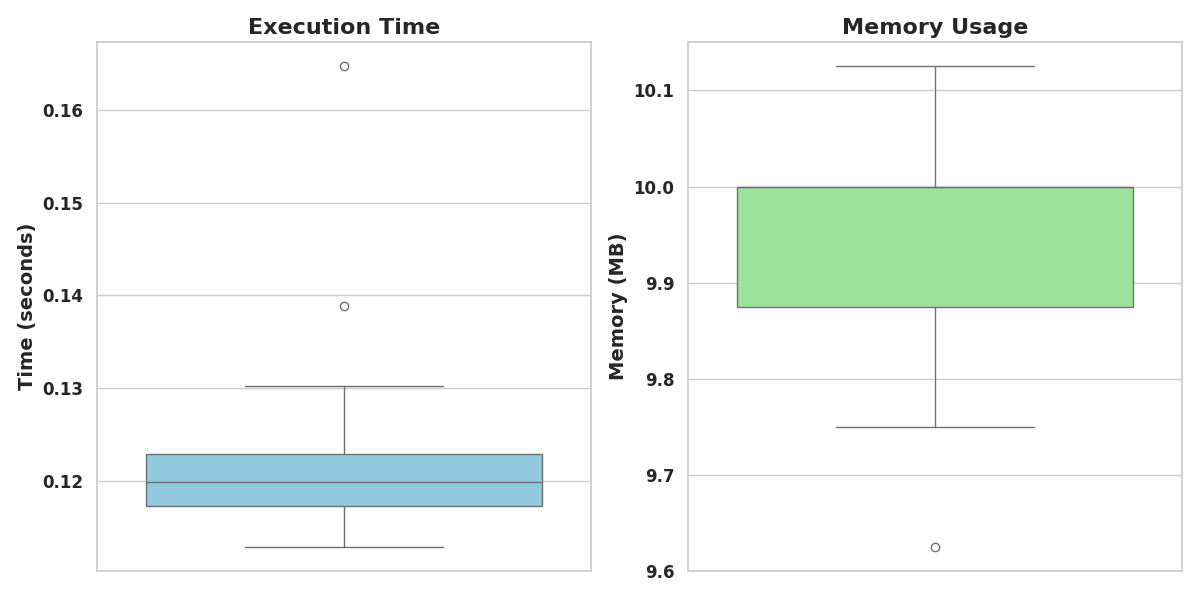}
        \caption{ZK proof verification requirements.}
    \end{subfigure}
    \caption{Time and memory overhead of the included zkSNARK proof. The circuit consists of circa 600k R1 constraints.}
    \label{fig:implementation:zkbenchmark}
    %\vspace{0.4cm}
\end{figure*}

We measured total wall-clock time and peak resident memory of 100 zkSNARK generation and 100 verification runs.
The statistically analyzed results are depicted in \autoref{fig:implementation:zkbenchmark}.
Hence, these proofs can be generated even on commodity hardware.
%Additional storage requirements also include storing (on the blockchain) at least $32B$ of ID-hash and metadata per user with a verified identity.
%Moreover, the potential addition of zkSNARKs to verify social media for content creators would increase the requirements to another $288B$ per social media proof.

\subsubsection{\textbf{Scaling Function Application}}

We limit creator participation in consensus (see \autoref{sec:design:bstrap:initialization}) because computing individual scaled-down effective social capital on the fly introduces non-negligible overhead.
Assuming 120 million YouTube channels~\cite{Broz2025YouTubeChannelsStatistics} representing content creators, this global calculation takes $\sim1.5$s serially or $\sim0.3$s in parallel.
Given a $12$s block-production timeslot for each round~\cite{ETHwhitepaper}, such overhead could noticeably hinder regular operation and safety margins.
By introducing a \textit{consensus participation threshold} of 10,000 full endorsements, only 3 million channels remain (a $97.5\%$ reduction)~\cite{Broz2025YouTubeChannelsStatistics}.
This lowers the per-slot serial computation time to about $0.03$s and the parallel time to $0.005$s in practice (using 8 CPU threads).
\section{Security Analysis}
\label{sec:security-analysis}

%This section analyzes potential attack vectors that could be exploited within our system, considering its specific intricacies.
%The primary focus of this section is on the threats specific to our proposed design. % and the concept of social capital used as a staking resource.
%We recall that credential issuers are considered trusted (see \autoref{sec:problem}), and we will not deal with attacks related to them.

\subsubsection{\textbf{Sybil Identities (Attacker 1)}}
\label{sec:security-analysis:attacks:IDP}
The goal of this attacker is to create a large number of Sybil accounts to artificially increase her social capital.
%\ih{Treba dorobit ked bude vyrieseny uniqueness.}
%
%Suppose there is an attacker, wishing to create numerous Sybil accounts to increase their influence.
%
The attacker would attempt to forge ID-hashes by using nonexistent data.
However, as we require a supplementary zkSNARK proof of a correct ID-hash creation, and we consider zkSNARKs sound and true zero knowledge (assumptions from \autoref{sec:problem}), this vector is not feasible.
Moreover, the VC contains a signature of the issuer, which cannot be forged without the knowledge of its private key. % it is not feasible to perform this attack.

\subsubsection{\textbf{Centralization of Social Capital (Attacker 2)}}
\label{sec:security-analysis:social-capital:monopoly}
The goal of this attacker is to dominate the system by holding an excessive amount of social capital and thus to produce numerous blocks (potentially subject to MEV profits).
Although the goal of this attacker is not to break safety by acting adversarially when exceeding the 33\% threshold of LMD-GHOST~\cite{gasper}, we also mitigate this problem.
In particular, effective social capital applies a non-linear scaling function (i.e., $log_2() \text{ or } sqrt()$) that compresses excessive influence of consensus nodes: see example of two consensus power distributions in \autoref{tab:security-analysis:social-capital:monopoly}, where the exceeded liveness-guarantee threshold is depicted in red.
Note that the maximum amount of active social capital, while not exceeding this threshold, is $\approx87\%$ and $\approx49\%$, when using $log_2()$ and $sqrt()$ scaling, respectively.
For comparison, we also showed two setups of quadratic voting  (w. and w/o. splitting votes), where it can be seen that it does not mitigate this problem since it scales down the social capital before assignment, keeping the ratios of social capital the same as before scaling.

% (a practical example can be seen in \autoref{tab:security-analysis:social-capital:monopoly}). 
% This method retains a performance edge for reputable validators but imposes diminishing returns on accumulated social capital, thereby:
% \begin{compactitem}
% 	\item Preventing oligarchic control over block production.
% 	\item Preserving profitability of smaller content creators.
% 	\item Maintaining equitable consensus distribution and long-term network health.
% \end{compactitem}

\begin{table}[t]
    \centering
    \scriptsize
    \setlength{\tabcolsep}{3pt}
    \caption{Scaling function under various distributions of consensus power. %\ih{Evaluation: Note that we assumed active social capital only for a comparison in the case of quadratic voting experiments, while quadratic voting does not consider it.}
    }
    \vspace{0.1cm}
    \label{tab:security-analysis:social-capital:monopoly}
    \begin{subtable}{1\columnwidth}
        \centering        
        \begin{tabular}{ r c  c  c  c  c }
             \toprule
             ~ & \textbf{Node 1} & \textbf{Node 2} & \textbf{Node 3} & \textbf{Node 4} & \textbf{Node 5} \\  
             \cmidrule(l){2-6}\cmidrule(r){6-6}
            \textbf{Active social capital} & \textcolor{red}{\textbf{40\%}} & 25\% & 15\% & 12\% & 8\% \\
            \midrule
            \textbf{q-voting} & \textcolor{red}{\textbf{40\%}} & 25\% & 15\% & 12\% & 8\% \\
            % q-voting (10 voters) & 24\% & 21\% & 21\% & 17\% & 17\% \\
            \textbf{q-voting (w. split vote)} & \textcolor{red}{\textbf{40\%}} & 25\% & 15\% & 12\% & 8\% \\
            \textbf{$\mathbf{sqrt()}$ scaling} & 29.43\% & 23.27\% & 18.02\% & 16.12\% & 13.16\% \\
            \textbf{$\mathbf{log_2()}$ scaling} & 23.32\% & 21.49\% & 19.5\% & 18.63\% & 17.06\% \\
            \bottomrule
        \end{tabular}
        %\vspace{-0.35cm}
        \caption{Mild monopolization.}
    \end{subtable}

    \begin{subtable}{1\columnwidth}
        \centering        
        \begin{tabular}{ r c  c  c  c  c }
             \toprule
             & \textbf{Node 1} & \textbf{Node 2} & \textbf{Node 3} & \textbf{Node 4} & \textbf{Node 5} \\            
             \cmidrule(l){2-6}\cmidrule(r){6-6}
            \textbf{Active social capital} & \textcolor{red}{\textbf{70\%}} & 12\% & 8\% & 5\% & 4\% \\
            \midrule
            \textbf{$\mathbf{sqrt()}$ scaling} & \textcolor{red}{\textbf{44.7\%}} & 20.69\% & 15.11\% & 11.94\% & 7.56\% \\
            \textbf{$\mathbf{log_2()}$ scaling} & 28.67\% & 21.93\% & 19.18\% & 17.12\% & 13.11\% \\
            \bottomrule
        \end{tabular}
        %\vspace{-0.35cm}
        \caption{Strong monopolization.}
    \end{subtable}

    % \begin{subtable}{1\columnwidth}
    %     \centering        
    %     \begin{tabular}{ r c | c | c | c | c }
    %          & Node 1 & Node 2 & Node 3 & Node 4 & Node 5 \\
    %         \toprule
    %         Active social capital & 90\% & 5\% & 3\% & 1.5\% & 0.5\% \\
    %         \midrule
    %         $sqrt()$ scaling & 64\% & 15\% & 11\% & 6\% & 4\% \\
    %         $log_2()$ scaling & 39\% & 22\% & 17\% & 13\% & 9\% \\
    %         \bottomrule
    %     \end{tabular}
    %     \caption{Extreme monopolization.}
    % \end{subtable}
    \vspace{-2em}
\end{table}

\subsubsection{\textbf{Extracting Private Data from the Blockchain (Attacker 3)}}
\label{sec:security-analysis:zkps}
zkSNARK-based user registration hides user identities on-chain, so attackers cannot extract personal data. 
However, zkSNARKs require a \textit{trusted setup} that generates secret \textit{toxic waste} which must be destroyed, since leakage would allow an adversary to forge proofs and break soundness. 
This risk could be mitigated by multi-party computation, assuming at least one participant is honest. 
As an alternative, zkSTARKs~\cite{ben‑sasson2018} avoid trusted setup and are post-quantum secure, at the cost of larger proofs and longer proving times.

Another way to reveal personal data is by brute-forcing ID-hashes, especially when the attacker knows the victim's identity information.
This is largely mitigated by including the VC identifier in the ID-hash, which adds 32B of entropy.
However, if a user re-issues VCs (e.g., after compromising $SK_B^u$), the trusted issuer must reuse the same VC identifier; otherwise, the user could create Sybil accounts. 
Alternatively, issuers can use a TEE-based secure logging approach~\cite{paccagnella2020custos,homoliak2020aquareum} that publicly tracks all VCs issued for each user to increase trustworthiness.

\subsubsection{\textbf{Mapping IPs to Blockchain Addresses (Attacker 3)}}
\label{sec:security-analysis:ipmapping}

%Linking IP to blockchain addresses poses a deanonymization risk.
Even though addresses are pseudonymous, transactions or network behavior might allow %observers, wishing to deanonymize creators,
adversaries to correlate blockchain addresses with IP addresses, especially if creators do not employ obfuscation tools.
This risk is increased for content creators, whose on-chain identities are public (by their choice). % in order to receive endorsements. 
It may expose them to ``doxxing'' (i.e., publicly revealing their private information without consent, often with malicious intent) or physical threats.
An optional mitigation for users is to use anonymization tools like VPNs, Tor, JonDonym, etc.

\iffalse
\subsubsection{\textbf{DoS on the Leader (Attacker 4)}}
\label{sec:security-analysis:consensus:election}

Ethereum’s use of Randao for on-chain randomness enables deterministic prediction of validator assignments within an epoch~\cite{LeastAuthority2018Eth2Audit,whisk}.
As randomness is derived from BLS-signed data and publicly known ahead of time, attackers can identify consensus nodes and launch DoS attacks on upcoming leaders, preventing them from proposing blocks and leading to penalties. % and missed rewards.
%
A mitigating solution is Whisk-SSLE~\cite{whisk}, a privacy-preserving leader election scheme that obfuscates proposer identities using anonymous registration and only reveals the leader at the time of block proposal, significantly reducing DoS feasibility.
%Our design is compatible with Whisk-SSLE. % of which it forms a part.
\fi

\subsubsection{\textbf{Off-chain Attacks on Social Capital}}
\label{sec:security-analysis:social-capital}

Content creators may exert off-chain influence over users via different non-explicit mechanisms that can manipulate endorsements without direct transactions.
For example:
\begin{compactitem}
	\item \textbf{Social pressure:} Users may feel compelled to support creators to avoid FOMO or to gain social standing.
	\item \textbf{Threats:} In cases where users reveal their identity, threats or harassment could coerce unwanted behavior.
\end{compactitem}
Such off-chain behavior remains undetectable on-chain.
However, a countermeasure might already exist within the system, stemming from the underlying nature of social media: re-assigning passive social capital.
We propose a system where an endorsement can be moved to a different creator, as the user could reassign their social capital away from the briber upon receiving a reward or when the coercive threat subsides.
This turns potential plutocratic dominance from a one-time investment (in a monetary system) to a continuous expense due to the trivial reassignment of social capital.

\section{Discussion}

\label{sec:discussion}

% This section highlights limitations, open questions, and possible extensions of the proposed system.

%\ih{Add delegated staking -- problemy co to moze sposobit (centralizacia) a kedze je to existujuci problem Etherea my sa tym dalej nezaoberame.}

\subsubsection{\textbf{Effects of Penalization}}
\label{sec:discussion:slashing}
%\ih{TODO}
One might argue that the effect of penalization (see \autoref{sec:design:overview:incentives:penality}) might be different in a PoSC system than in PoS, where crypto-tokens are staked instead of social capital.
If we were to consider that a (penalized) validator has the ability to migrate to another PoSC platform to escape the consequences of slashing rules, we would also have to consider the complexity of moving their followers to that platform.
We, however, work under the assumption that PoSC platforms are independent, i.e., social capital is a non-transferable resource even in the case of cross-chain communication.

\subsubsection{\textbf{Delegation of Social Capital}}
\label{sec:discussion:trnasfer-sc}

Delegating social capital would also allow users to redirect it to other creators they believe in, turning social capital into a transferable resource.
However, a high-influence creator could circumvent the diminishing returns of the scaling function by delegating social capital to a set of puppet accounts.
To mitigate this, we could restrict delegation to effective social capital; thus, redistributing influence would offer no advantage.
However, tracking and managing delegated social capital, particularly across multiple nested delegations, would require significant overhead.
If a content creator's social capital is partially or wholly delegated, it is unclear whether past endorsements made by their followers retain their original weight.%\ih{skratit}

\subsubsection{\textbf{Hybrid Combination with Stake}}
\label{sec:discussion:hybrid}

A promising extension is a hybrid model combining social capital with monetary stake to secure consensus.  
Validators would stake both forms, with influence derived from their aggregate contribution.
This approach leverages two dimensions of trust: economic security from financial risk and reputational legitimacy from community support.  
A key issue is defining the exchange rate between social and monetary stake.  
Parity (e.g., one-to-one) might be arbitrary and thus risk bias of one group.  
Alternatively, influence could be dynamically weighted based on uptime, reputation, or external oracle data, which opens avenues for manipulation, regulation, or market distortion.

\subsubsection{\textbf{Alignment of Incentives for Followers}}
It may appear that a PoSC system will naturally gravitate towards a state where followers cluster around a creator who offers the most value in return for an endorsement, in contrast to genuinely assigning social capital (disregarding bribes or coercion \autoref{sec:security-analysis:social-capital}).
For example, some creators might offer more attractive or limited benefits, such as exclusive or early access content, personalized services, or unique experiences.
Such incentive structure naturally encourages creators to compete on genuine value creation.
% Followers might tend to gravitate toward creators who offer the most attractive or limited benefits, such as exclusive content, early access, personalized services, community status, or unique experiences.
This competition closely mirrors creator–fan economies on existing platforms (e.g., Patreon exclusives~\cite{patreon} or YouTube Channel Memberships~\cite{youtube2}), but in a more transparent and verifiable setting.
%\review{It seems like such a system will naturally gravitate towards whoever offers users the most value in return (bribes, offers, services, etc.) for their vote.}

\subsubsection{\textbf{Financial Incentives for Followers vs. Decentralization}}
So far, we assumed that incentives for followers were in the form of services. 
However, let us discuss another hypothetical setting.
If we were to assume that incentives for followers are purely financial, such as a direct token reward for receiving (e.g., marketing) content or interacting with it might have a self-balancing feedback effect on the distribution of consensus power. 
In particular, content creators would compete to achieve a maximum number of followers whose social capital contribution they are willing to reward, while still remaining profitable.
%In particular, content creators would, on one hand, compete for a maximum number of followers contributing to their active social capital, but on the other hand, they must financially reward these followers and still remain profitable -- 
This is further influenced by the scaling of effective social capital that grows only logarithmically with the number of followers. %(\ih{autoref on new a graph}).
%Therefore, the followers might prefer to distribute their passive social capital among as much as possible independent content creators to maximize their profits.
In this way, the system naturally enforces decentralization. 

\subsubsection{\textbf{Endorsement Based on Different Aspects of Social Capital}}
% \ih{todo}
One concern is that endorsers may not award their social capital purely based on merit, quality, or rewards associated with the creator, but also on other aspects, such as religion, nationality, gender, etc.
Nevertheless, this is not a novel problem for our design or the existing web-based social networks.
Endorsers, similarly to followers or subscribers on Web2 platforms, are encouraged to form like-minded communities.
While such clustering can reinforce so-called echo chambers, it also reflects genuine user preferences for trusted or culturally resonant voices.

\iffalse
\subsubsection{\textbf{Effective SC Stake vs. Consensus Threshold}}

\jm{
In conducted experiments, a metric of the number of slots required to reclaim the same amount of tokens lost for social capital gain clearly shows that the most efficient number of active social capital exactly matches the consensus participation threshold. 
Nevertheless, this promotes decentralization...}

\subsubsection{\textbf{Game Theory Surrounding SC Reassignment Transactions}}

\jm{
A situation could arise in which block producers are willing to process the social capital assignment transactions for creators if their active social capital post-assignment is lower or higher than the participation threshold.
However, transactions that would make a creator exceed the threshold by processing the transaction and thus lowering the chances of other creators being selected as block producers could cause these types of transactions to stagnate.
}
\fi

\subsubsection{\textbf{Post-Quantum Cryptography}}
\label{sec:discussion:pqc}
Since zkSNARKs, BLS (as well as ECDSA) signatures are not post-quantum secure (as mentioned in \autoref{sec:security-analysis:zkps}), another extension of the protocol can be the utilization of zk-STARKs~\cite{ben‑sasson2018} and lattice-based signing schemes~\cite{nist2024fips204,ducas2017dilithium}.
Nevertheless, the problem of post-quantum secure signature aggregation remains. %\ih{over + dodaj nejaku citaciu z ETH, pripadne ako chcu riesit oni.} and is one of the focal points of Ethereum's scientific debate.
Possible post-quantum algorithms involve Falcon signatures with LaBRADOR aggregator~\cite{nevado2025lattice}, or zkSTARKs proving the validity of many hash-based signatures.

\subsubsection{\textbf{Trusted Credential Issuer}}

The protocol currently assumes that credentials signed by recognized legal authorities are trustworthy.
While feasible in the short term, this assumption overlooks the evolving nature of trust: trusted issuers may become compromised and potentially issue fraudulent credentials.
We mitigate this by revocation of trusted issuers by supermajority consensus.
\section{Related Work}
\label{sec:related-work}
%\jm{Check the Basic Attention Token}

% The study of social capital in digital ecosystems has gained traction as platforms increasingly leverage human capital, attention, and influence for economic gain.
% Research highlights the interplay between centralized and decentralized platforms, each offering distinct mechanisms for measuring and monetizing social capital.
% Centralized systems rely on algorithmic control, while decentralized platforms emphasize user autonomy and transparency through blockchain technology.
%
This section reviews key platforms and their approaches to social capital, focusing on their governance, consensus mechanisms, monetization models, and implications for user control and equity.

\subsection{Identity \& Uniqueness on Blockchains}

\textbf{Worldcoin}~\cite{WorldCoinWhitepaper} aims to provide a global digital identity system grounded in \textit{Proof-of-Personhood} and \textit{Proof-of-Uniqueness}~\cite{PoPidea} principles. It enforces one-identity-per-user through iris biometrics captured by a proprietary \textit{Orb} device. Such guarantees could, in principle, serve as an external source of uniqueness for our protocol.
\textbf{BrightID}~\cite{brightidWhitepaper} offers a decentralized proof-of-uniqueness scheme that avoids personal identifiers and instead evaluates a user-generated social graph.
Network-structure analysis~\cite{wang2020structurebasedsybildetectionsocial,breuer2020friendfauxgraphbasedearly,furutani2022interpretinggraphbasedsybildetection} is used to differentiate authentic users from Sybil clusters, although the strength of these guarantees depends on both graph integrity and detection quality.
\textbf{Proof of Humanity (PoH)}~\cite{PoH2021} maintains an Ethereum-based registry of verified identities.
Users provide a photo, a video statement, and a vouch from existing members, after which entries remain open to community challenges. %resolved through the \textit{Kleros} arbitration system.
Its effectiveness relies on sustained vigilance, so in a large registry, undetected duplicates become increasingly plausible, leaving room for Sybil attacks.

\subsection{Social Media}

The PoSC protocol, when deployed in an application, resembles a Web2 social platform by rewarding creators for engagement and establishing a social interaction layer.
\textbf{Centralized Platforms}~\cite{youtube,youtube2,instagramTiktok,onlyfans,patreon} convert engagement metrics into social capital but remain governed by corporate algorithms.
They follow either an advertisement model (YouTube, TikTok, Instagram) or a subscription model based on direct support (OnlyFans, Patreon).
Creators remain exposed to algorithmic volatility, moderation policies, deplatforming, and revenue distribution concerns~\cite{Hödl2023,Hua_2022}.
\textbf{Decentralized Platforms:} Several blockchain applications incorporate social capital at the application layer.
\textbf{Steemit}~\cite{steemwp} uses a dual-token system where engagement is rewarded from a \texttt{reward pool} weighted by \texttt{Steem Power}, although this design has been criticized for empowering large stakeholders~\cite{li2021incentivizedblockchainbasedsocialmedia}.  
\textbf{Friend.tech}~\cite{friendtech} introduces tradeable ``keys'' priced by a quadratic bonding curve to unlock content, which can shift attention toward speculative key trading.
\textbf{Farcaster}~\cite{farcaster,farcaster2} offers decentralized identities stored on-chain with application data hosted on a data-availability network and does not natively embed monetization.
None of these systems integrates social capital into the consensus layer, which is the focus of our work.

% While all platforms aim to reduce reliance on centralized intermediaries, they grapple with challenges, such as incentive misalignment, wealth concentration, speculative behavior, and scalability.
% Nonetheless, they demonstrate that social capital can be formalized, incentivized, and transacted in decentralized systems.
\section{Conclusion}
\label{sec:conclusion}

In this paper, we proposed a new Proof-of-Social-Capital consensus protocol, a protocol that shifts the scarce resource from a financial one to social influence, while remaining secure against legacy threats.
To achieve these properties, we built upon Ethereum's proven Proof-of-Stake protocol.
We employed additional features, such as identity verification using zero-knowledge proofs for Sybil protection, a scaling function for social capital to prevent monopolization, and user incentives to encourage adoption.
We created a proof-of-concept implementation to demonstrate the feasibility of such a protocol, laying the groundwork for future exploration of socially driven consensus systems in decentralized environments.

\bibliographystyle{IEEEtran}
\bibliography{ref}

\end{document}